\documentclass[a4paper, 11pt]{article}

\usepackage[german, english]{babel}
\usepackage{authblk}
\usepackage{setspace}
\usepackage{array}
\usepackage{tabularx}
\usepackage{graphicx}
\usepackage{amsmath}
\usepackage[left=2.0cm,right=2.0cm,top=2.0cm,bottom=2.0cm]{geometry}
\usepackage{float}
\usepackage[authoryear,round]{natbib}
\usepackage{url}
\usepackage{doi}
\usepackage{xcolor}

\newcommand\tophline{\hline\noalign{\vspace{1mm}}}
\newcommand\middlehline{\noalign{\vspace{1mm}}\hline\noalign{\vspace{1mm}}}
\newcommand\bottomhline{\noalign{\vspace{1mm}}\hline}
\newcommand\authorcontribution{\endlist\egroup}
\newcommand\competinginterests{\endlist\egroup}
\newcommand\acknowledgments{\endlist\egroup}

\begin{document}

\doublespacing

\title{The global land water storage data set release 2 (GLWS2.0) derived via assimilating GRACE and GRACE-FO data into a global hydrological model}
\author[1]{Helena Gerdener}
\author[1]{Jürgen Kusche}
\author[1]{Kerstin Schulze}
\author[2,3]{Petra Döll}
\author[4]{Anna Klos}

\affil[1]{Institute of Geodesy and Geoinformation (IGG), University of Bonn, Bonn, Germany.}
\affil[2]{Institute of Physical Geography, Goethe University Frankfurt, Frankfurt am Main, Germany}
\affil[3]{Senckenberg Leibniz Biodversity and Climate Research Centre Frankfurt (SBiK-F), Frankfurt am Main, Germany}
\affil[4]{Faculty of Civil Engineering and Geodesy, Military University of Technology, Warsaw, Poland.}

\affil[ ]{Corresponding author: Helena Gerdener, gerdener@geod.uni-bonn.de, Nussallee 17, 53115, Bonn, Germany.}

\date{ }

\maketitle

\begin{abstract}
We describe the new global land water storage data set GLWS2.0, which contains total water storage anomalies (TWSA) over the global land except for Greenland and Antarctica with a spatial resolution of 0.5$^\circ$, covering the time frame 2003 to 2019 without gaps, and including uncertainty quantification. GLWS2.0 was derived by assimilating monthly GRACE/-FO mass change maps into the WaterGAP global hydrology  model via the Ensemble Kalman filter, taking data and model uncertainty into account. TWSA in GLWS2.0 is then accumulated over several hydrological storage variables. In this article, we describe the methods and data sets that went into GLWS2.0, how it compares to GRACE/-FO data in terms of representing TWSA trends, seasonal signals, and extremes, as well as its validation via comparing to GNSS-derived vertical loading and its comparison with the NASA Catchment Land Surface Model GRACE Data Assimilation (CLSM-DA). We find that, in the global average over more than 1000 stations, GLWS2.0 fits better than GRACE/-FO to GNSS observations of vertical loading at short-term, seasonal, and long-term temporal bands. While some differences exist, overall GLWS2.0 agrees quite well with CLSM-DA in terms of TWSA trends and annual amplitudes and phases.
\end{abstract}

\section*{Keywords}
GLWS, GRACE, GRACE-FO, total water storage anomalies, data assimilation, global 

\section*{Highlights}
\begin{itemize}
    \item We describe the new global land water storage data set GLWS2.0, which contains total water storage anomalies over the global land with a spatial resolution of 0.5$^\circ$, covering the period 2003 to 2019 without gaps, and including uncertainty quantification.
    \item GLWS2.0 synthesizes monthly GRACE/-FO mass change maps with daily precipitation and radiation data via the WaterGAP model framework, taking data and model uncertainty into account.
    \item Here we describe the methods and data sets that went into GLWS2.0 and its validation from a geodesy applications perspective. We find that, in the global average, GLWS2.0 fits better than GRACE/-FO to GNSS observations of vertical loading.
\end{itemize}

\newpage
\section{Introduction}
The GRACE and GRACE-FO satellite missions \citep{tapley_contributions_2019} have provided unique and unprecedented data sets for studying the terrestrial water cycle, its natural variability, and its response to radiative forcing, land use change, and direct anthropogenic alterations \citep{rodell2018}. GRACE/-FO derived maps of monthly total water storage anomalies (TWSA) have been used to quantify long-term trends in glacier mass balance \citep{wouters_global_2019}, groundwater stress \citep{richey_quantifying_2015}, large-scale droughts \citep{zhao_global_2017, gerdener_framework_2020} and floods \citep{reager_river_2014, han_grace_2021}, the vegetation response to moisture \citep{a_satellite-observed_2017, gerdener2022}, to evaluate land surface and atmospheric models at different timescales \citep{springer_new_2014}, and to improve the representation of e.g. soil processes in models \citep{swenson_assessing_2014}.

Apart from these climate and hydrology applications, GRACE/-FO data are increasingly employed in geodesy and geodynamics in lieu of model-based mass redistribution maps, e.g. for deriving environmental loading computations for satellite altimetry \citep{ray_monthly_2013} and GNSS \citep{chanard_toward_2018, memin_correcting_2020}, deriving geocenter estimates \citep{wu_new_2017}, simulating the hydrological (HAM) and ocean (OAM) angular momentum contributions to Earth rotation variations \citep{seoane_use_2009,jin_hydrological_2010}, inferring rheological properties of the Earth \citep{broerse_postseismic_2015, rovira-navarro_grace_2020}, or for deriving empirical or semi-empirical models of glacial isostatic adjustment GIA \citep{sun_global_2020}.

Yet the effective GRACE/-FO spatial resolution of about 300km, after appropriate post-processing, is still too coarse for several applications in hydrology and climate but also in geodesy and geodynamics (\cite{pail_science_2015, wiese_mass_2022}. In addition, GRACE errors are not stationary with respect to time due to switching off the thermal control of the accelerometers in 2010 and other hardware problems. Monthly data gaps exist, and an eleven-month gap between the two missions could not be avoided. All this means that the time series of TWSA maps are difficult to work with if either spatial resolution or temporal consistency is an issue. Several simulation studies have shown that improved mission concepts with multiple spacecraft and upcoming technology are expected to lead to notable improvements, but not to a quantum leap in spatial resolution (see \cite{pail_science_2015, wiese_mass_2022} and references therein).

In recent years, studies made efforts to predict unobserved TWSA (e.g. during GRACE gaps or prior to the GRACE era) by combining GRACE/-FO maps with remote sensing or in-situ data via statistical or machine learning approaches \citep{humphrey_grace-rec_2019, li_long-term_2021}. Also, many projects apply formal data assimilation (DA) techniques \citep{zaitchik_assimilation_2008, tangdamrongsub_data_2015, girotto_assimilation_2016, li_global_2019, springer_evidence_2019} in order to disaggregate observed TWSA into changes e.g. in the groundwater, soil, snow or surface water storages. But to our knowledge, they all focused on hydrological applications, and only a few (if any) data products are publicly available including quantification of their uncertainty. For example, \cite{tangdamrongsub_data_2015} and \cite{tangdamrongsub_multivariate_2020} investigate the improvement via assimilation in individual storages and fluxes by comparisons with independent groundwater, soil moisture, and streamflow records, and \cite{schumacher_systematic_2016} and \cite{khaki_assessing_2017} tested different assimilation schemes, but all these studies are regionally limited. Groundwater and soil moisture conditions from GRACE/-FO TWSA assimilation into the CSLM model \citep{li_global_2019} for the contiguous U.S.A. and for the global land surface can be downloaded from the National Drought Mitigation Center, University of Nebraska-Lincoln, archive (\texttt{https://nasagrace.unl.edu/}) in form of indicator percentile maps, but the underlying water storage grids are not publicly available. Lastly, several assimilation projects make use of GRACE/-FO mascon (i.e. level-3) data grids - while we acknowledge that these are easy to use and competitive to the standard GRACE/-FO solutions for many applications, the use of level-2 GRACE/-FO data allows one to design more flexible and optimized analysis schemes since the smoothing within the mascon processing cannot be undone in post-processing.

Following methods outlined in \citep{eicker2014}, we developed a global data assimilation framework based on the Ensemble Kalman filter approach \citep{evensen_ensemble_2003} to produce a global land water storage (GLWS) data  set with 0.5$^\circ$ spatial resolution, here in version 2.0 (GLWS2.0). Via the WaterGAP \citep{muller_schmied_global_2021} modeling framework, TWSA is at the same time vertically disaggregated into ten different water compartments including groundwater, soil, snow, and surface water (river, reservoir, wetland, and lake) storage changes. WaterGAP runs had been evaluated before with respect to GRACE/-FO TWSA data, both at grid and basin scale and at various timescales, and were found to perform with mixed results \citep{scanlon_global_2018, scanlon_tracking_2019, muller_schmied_global_2021}. Similarly, GRACE TWSA maps had also been used to calibrate WaterGAP model parameters before \citep{werth_calibration_2010}.

GLWS2.0 can be viewed as synthesized or merged from GRACE/-FO mass change grids and precipitation and radiation data grids (the WaterGAP forcing data), taking data uncertainties into account, by including several flow-storage relations that are conceptualized in the hydrological model. In other words, with GLWS2.0 we provide a 'level-3', downscaled TWSA data set (next to 'level-2' hydrological variables, such as groundwater and surface water storage anomalies) that we suggest can provide reference surface mass change maps for geodesy and geodynamic studies. Other studies \citep{eicker2014, schumacher2018, gerdener2022} had assimilated TWSA in WaterGAP but this is the first global assimilation with this model. As mentioned above, several others have assimilated TWSA data into hydrological or land surface models, but many early studies used basin-averaged in place of gridded TWSA data, the resulting downscaled maps are mostly not publicly available, and the models that were used employed very different representations of the terrestrial water storages and fluxes (e.g. unlike WaterGAP most models miss anthropogenic water withdrawals).

We would like to point out that the new 'WaterGAP reanalysis' GLWS2.0 could be used as well in hydrology and climate modeling and validation exercises, having significantly higher effective spatial resolution than GRACE/-FO, but with the caveat that it is not a pure observational data set. We show in this contribution that GLWS2.0, in the spectral (spherical harmonic) domain, provides a smooth transition in power between GRACE data and WaterGAP simulation data. An earlier version of this data set (GLWS1.0) had been distributed within the GRACE community before; significant updates are a new WaterGAP model version with an improved representation of surface water and groundwater abstraction and several bug fixes.

This contribution focuses on the monthly GLWS2.0 TWSA data sets, how they were generated, and how they compare to the GRACE/-FO data and to the WaterGAP data. For an evaluation against independent data, we predict monthly-average vertical station movements from TWSA grids for about 1030 IGS sites, and compare to the Repro3 solution \citep{rebischung_terrestrial_2021}; i.e. to vertical motion as derived from GPS, GLONASS and Galileo observations. The first applications of GLWS in hydrological research were described in \cite{gerdener2022} and \cite{dibi2022}.

This note is organized as follows: First, we introduce the gridded GRACE and GRACE-FO data that we use for assimilation and to which we later compare, as well as the WaterGAP model, forcing data, and the GNSS data that we finally use for evaluation. Then, our data assimilation framework is described.  Next, we compare GLWS2.0 to GRACE/-FO and WaterGAP TWSA maps in the spatial as well as in the spectral domain, we will provide an uncertainty quantification, and we compare to another global assimilation-derived TWSA data set. Finally, we compare to GNSS observations and provide some conclusions and an outlook.

\section{Data and models}
\subsection{GRACE}
\label{sec_grace}
The Gravity Recovery and Climate Experiment (GRACE) satellite mission by the German aerospace center (DLR) and NASA's Jet Propulsion Laboratory (NASA/JPL) carried a K-band instrument to perform range-rate measurements between two co-orbiting GNSS-tracked platforms. GRACE was launched in 2002; although it was planned as a five-year mission it ended in 2017 after nearly 15 years of operation. Its successor GRACE-FollowOn (-FO) was launched in 2018 and is still in orbit.

In this study, we use level-2 spherical harmonic geopotential coefficients derived from GRACE/-FO instrument data at TU Graz (ITSG 2018 and operational, \cite{kvas2019itsg}; \cite{mayer2018itsg}). We chose the ITSG data product because it is accompanied by full error information -- i.e. the normal equations -- which we deem important to represent the dominating uncertainty patterns of GRACE in the assimilation. Level 2 data are then further converted to TWSA maps following the standard processing steps (see \cite{gerdener2022}): certain lower degree coefficients are replaced, we apply a de-striping procedure \citep[DDK3, ][]{kusche_approximate_2007}, remove the temporal mean from 2003 to 2016 (time span was decided before the GRACE/-FO gap), and finally synthesize TWSA maps following \cite{wahr_time_1998}. Gravity changes related to glacial isostatic adjustment are removed using the \cite{a2013} model.

In the standard GRACE/-FO post-processing at IGG/University of Bonn, we generate TWSA maps at 0.5$^\circ$ resolution, consistent with the spatial resolution of most global hydrology models. However, \cite{schumacher_systematic_2016} compared various spatial aggregation scales applied to TWSA maps, when including them into the assimilation, and found that  0.5$^\circ$ and even 1.0$^\circ$ can lead to numerical instabilities in the variance-covariance matrix, which is used to weight observations and model simulations. Therefore, as a compromise, here we transform the SHC to a 4$^\circ$ grid prior to assimilation. This can be viewed as an aggregation to numerically regularize the estimation, but it clearly introduces fewer constraints than the use of basin-averaged TWSA.

As mentioned above, in order to represent the dominating error patterns of GRACE, i.e. the latitudinal dependency due to orbital coverage, we derive covariances from the level-2, fully populated normal equations, and propagate them through our post-processing until we arrive at a fully populated level-3 (TWSA grid) variance-covariance matrix. However, we found that remaining spatial error correlations for 4$^\circ$ TWSA maps can still induce numerical instability during the update step of the assimilation. This is why we finally concluded to suppress any spatial correlation at the grid level in the representation of GRACE errors within the assimilation. In other words, when vector ${\bf y}$ represents observed TWSA on $n$ grid points, the GRACE uncertainties are represented in the DA via 
    
\begin{equation}
\label{eq001}
{\bf C}_{\hat{\bf y}\hat{\bf y}} = 
\begin{bmatrix}
    \sigma_{y_1}^2 & 0 & 0 & ... & 0 \\
    0 & \sigma_{y_2}^2 & 0 & ... & 0 \\
    ... & ... & ... & ... & ...  \\
    0 & 0 & 0 & ... & \sigma_{y_n}^2 \\
\end{bmatrix}
\end{equation}

with the 1-sigmas $\sigma_{y_i}$ being, in first approximation, dependent on latitude.  We stress that suppressing spatial correlation in assimilation frameworks is a common procedure, and what we describe above can be viewed as a form of observation localization \citep{houtekamer_data_1998}. This is in line with simulation experiments by \cite{schumacher_systematic_2016} who found that suppressing spatial correlations, within a DA framework very similar to ours, did not negatively affect results on average.

Some studies have suggested assessing GRACE TWSA errors by comparing different GRACE solutions; either derived from level-2 data or from mascon processing. We refrain from this approach here since (1) these differences may not reflect real uncertainty since all processing centers use similar analysis schemes and the same input data; and (2) the GRACE uncertainties that we infer via rigorous error propagation may lack absolute calibration, they are likely more realistic in terms of spatial patterns. We also stress that for the assimilation here we could in fact ingest any GRACE solution together with any reasonable error covariance; since it is not straightforward to assume the different GRACE solutions have vastly different spatial error patterns.

\subsection{WaterGAP Global Hydrology Model and precipitation data}
In this study, we make use of the WaterGAP global hydrological model \citep{muller_schmied_global_2021}, version 2.2e. WaterGAP simulates water fluxes and storages on an 0.5$^\circ$ grid by solving vertical and horizontal water balance equations, representing ten different storage types in total: local lakes, global lakes, local wetland, global wetland, river, reservoirs, groundwater, soil moisture, snow and canopy storage. Unlike many land surface models, WaterGAP accounts for anthropogenic water use for irrigation, livestock farming, manufacturing, domestic use, and thermal power plant cooling. In its standard configuration, the model is calibrated against long-term annual river discharge, but these calibrated parameters are not used here. The current version implements, as a major update with respect to version 2.2d, an improved algorithm for surface and groundwater abstraction.

Hydrological models generally partition observed (or reanalysis, i.e. simulated but constrained by observations) precipitation into a variety of fluxes, some of which, like evapotranspiration, are again largely controlled by observations. This forcing data is, for our runs, derived from the GSWP3-W5E5 data set and it includes precipitation, temperature, longwave radiation, and shortwave radiation \citep{10.48364/ISIMIP.982724}.

For data assimilation we need to represent the uncertainty of the model states; i.e. for ten storage compartments on the global land, represented by the $0.5^\circ$-grid. In the EnKF this is achieved via an ensemble that accounts for perturbations of the initial model states, of the forcing data, and of 22 WaterGAP model parameters.

While WaterGAP appears as a fairly complete model in terms of processes that are represented conceptually, it is clear that certain mismatches must be expected when confronted with GRACE/-FO data. For example, glaciers are not represented as storage in the model
and real glacier mass imbalance observed by GRACE/-FO will lead to mismatch and potentially updating other storages in the model (e.g. snow storage or groundwater). Likewise,  anthropogenic water use is simulated via sectors models, but these rely on external data e.g. for irrigation which will be inevitably incomplete \citep[e.g.][]{muller_schmied_global_2021}. As a result, we expect that in heavily engineered regions the model deviates from GRACE, and TWSA in GLWS should be pulled towards what GRACE observes. Other potential sources of mismatch are reservoirs (WaterGAP mimicks reservoir operations in a very simplified form) and rivers (WaterGAP does consider river storage, but the river network topology itself is greatly simplified).

\subsection{GNSS validation data}
Even twenty years after the launch of GRACE, an independent assessment of the systematic and random uncertainty of satellite-derived land TWSA is difficult, due to the as-yet unprecedented nature of space-gravimetric data. Hydrological models are not sufficiently accurate, due to missing process representation but also uncertain forcing data, with spatially and temporally heterogeneous quality. `Test signals' have been frequently proposed, e.g. Sahara regions where low RMS is expected, and large lakes and reservoirs where water levels, bathymetry, and area change can be monitored independently; yet problems are suspected with groundwater flow in deep aquifers or reservoir seepage due to fractured geology. Measurements with superconducting gravimeters have been suggested for comparison to, and validation of, GRACE-derived gravity variations \citep{crossley_comparison_2012} but they are affected by local water table variations and other environmental factors, and their global coverage is quite limited.

Therefore, in this study we decided to compare GRACE/-FO and GLWS2.0 TWSA grids to vertical loading derived from hundreds of globally distributed GNSS sites, using methods that were described in \citep{blewitt_inversion_2003, wu_large-scale_2003, kusche_surface_2005}, and applied to evaluate GRACE solutions e.g. in \cite{chen_validation_2018}. GNSS loading evaluations rely on the validity of an underlying Earth model, which is usually assumed as radial-symmetric and purely elastic; real poroelastic or thermoelastic deformations are difficult to separate from observations and may mask elastic loading in certain geological settings \citep{carlson_seasonal_2020}, and the varying thickness of the lithosphere below topography and ocean is not taken into account. GNSS data are also affected by technique errors, however, with continuous reprocessing efforts the impact of these errors appears to be greatly reduced, and we believe the sheer number of displacement time series will eventually help to overcome the mentioned problems.

We use 1030 vertical displacement time series that were provided by the IGS as part of the Repro3 reprocessing \citep{rebischung_terrestrial_2021}. Repro3 includes data from the GPS, GLONASS, and Galileo constellations and it was generated as the GNSS contribution to the recent realization of the International Terrestrial Reference Frame \citep[ITRF2020, ][]{altamimi2021}. To be consistent with the TWSA maps, we removed non-tidal atmospheric and ocean loading effects from all time series and formed monthly averages. All time series (GNSS and predicted from GRACE/-FO and GLWS2.0) were detrended before comparison, in order to eliminate potentially incorrectly modeled glacial isostatic adjustment or local tectonic or anthropogenic subsidence and uplift. Finally, residual time series (i.e. explained variability) are split into short-term, seasonal, and long-term bands.

\section{Data assimilation framework}
\label{sec:Data assimilation framework}
The global land water storage data set version 2.0 (GLWS2.0) is produced by assimilating gridded GRACE and GRACE-FO-derived monthly TWSA maps into the WaterGAP global hydrological model, via an Ensemble Kalman Filter \citep[EnKF, ][]{evensen_ensemble_2003} approach as implemented in the Parallel Data Assimilation Framework \citep[PDAF, ][]{nerger2013}. The resulting data set represents thus an optimal synthesis of GRACE/-FO data and all data sets that went into the hydrological model. In other words, this synthesis seeks to fit observed TWSA grids (aggregated to $4^\circ$) within error bars, and at the same time, it solves the horizontal and vertical water balances as represented in the hydrological model (i.e. on a $0.5^\circ$ grid), again within error bars.

To this end, the uncertainty of the hydrological model simulation is represented via an $n$-member ensemble, where we take into account the uncertainty of forcing data as well as the uncertainty of some model calibration parameters. As a result, when no GRACE or GRACE-FO data are available, the GLWS data set represents the mean of an ensemble where each member is dynamically consistent with the model. It is important to recognize that this mean depends on the ensemble creation and thus will differ from published WaterGAP standard runs, even if there is no GRACE data within a particular month. It is also important to understand that the assimilation-derived GLWS TWSA data set does not represent a simple downscaling of the GRACE data, i.e. spatial smoothing of GLWS will not necessarily correspond to GRACE/-FO TWSA.

The monthly level-3 GLWS data represent the total water storage anomaly on $0.5^\circ$ grids; they are publicly available now from 01/2003 to 12/2019 and will be updated in the future. Additionally, the error standard deviation of GLWS TWSA at the grid cell level is provided, derived from the EnKF ensemble referred against the ensemble mean. The main updates with respect to GLWS1.0 were the use of an updated version of WaterGAP as described above, as well as minor bug fixes in the assimilation. Model initialization, spin-up period, and other details are as described in table \ref{table_DA_settings}. No model parameters were actually calibrated during this run. 
     
\begin{table*}[h]
\centering
\caption{\doublespacing Settings used in the data assimilation for the GLWS2.0 data set.}
\label{table_DA_settings}
\begin{tabular}{ | l | l | }
\tophline
Data assimilation/calibration method DA & GRACE-FO DA, no calibration\\
\middlehline
Time period & 01/2003 – 12/2019\\
\middlehline
Filter & EnKF\\
\middlehline
Forgetting factor & $1.0$\\
\middlehline
Ensemble size & $32$\\
\middlehline
Calibration parameter range & Small range\\
\middlehline
Initialization phase & 01/1995 – 12/2000\\
\middlehline
Spin up phase & 01/2001 – 12/2002\\
\middlehline
Open loop phase & 01/2003 – 12/2019\\
\middlehline
Temporal mean & derived from open loop 01/2003 – 12/2016\\
\bottomhline
\end{tabular}
\end{table*}

In the EnKF it is common to adjust the weighting between model prediction and data by 'inflating' the ensemble spread, either based on empirical considerations or on concepts very similar to variance component estimation. However, the inflation factor (the inverse of the 'forgetting' factor in table \ref{table_DA_settings}) was set to 1.0 for GLWS2.0., i.e. no weighting adjustment was applied (see Sec. \ref{sec_robustness}).

In order to simulate the combined effect of model structural errors and parameter uncertainty, for each of about 30 WaterGAP model parameters we define two alternative uncertainty bands, a 'small' (optimistic) and a 'large' (pessimistic) range. For each range,  minimum and maximum plausible parameter values were defined based on expert knowledge \citep[e.g. ][]{schumacher2018improving, werth2009integration, hosseini2020quantifying}, and we then assign triangular probability distributions to multiplier (i.e. scaling) parameters and uniform distributions to all other parameters. This allows us then to generate ensemble runs with model parameter perturbations. In this study, we used the small range.

In the EnKF, an ensemble of model simulations is run in parallel to represent simulation uncertainty, with the model states collected in ${\bf X}$. In a nutshell, once TWSA observations are assimilated, the entire ensemble of model states is updated to $\hat{\bf X}$  following
  
  \begin{equation}
    \label{eq002}
        \hat{\bf X} = {\bf X} + {\bf K}({\bf Y} - { H}({\bf X}))
    \end{equation}
  
where each column of ${\bf X}$ contains all model states for a single ensemble member, i.e. the entries for all ten water storages in all $x$ grid cells. Since we work with 32 ensemble members, ${\bf X}$ contains $n_e =32$ columns. In Eq. (\ref{eq002}), ${\bf K}={\bf W}{\bf W}'{\bf H}'({\bf H}{\bf W}{\bf W}'{\bf H} + {\bf G}{\bf G}')$ is the gain matrix whose task is to weight model and observation errors, which we both represent through ensembles (i.e. covariance matrices approximated through ${\bf W}{\bf W}'$ for simulated TWSA and ${\bf G}{\bf G}'$ for the observed TWSA). ${\bf Y}$ contains an ensemble of aggregated TWSA observations where perturbations have been sampled from the error maps Eq. (\ref{eq001}), and $H(\cdot)$ represents the (in our case linear, i.e. ${ H}({\bf X}) = {\bf H}{\bf X}$) operator that maps the model states to aggregated observations of total water storage variability \citep{eicker2014}.

The ensemble mean, for all ten storages in the model, is then provided by 

  \begin{equation}
    \label{eq003}
        \hat{\bf x} = \frac{1}{n_e}\hat{\bf X} \ {\bf 1}
  \end{equation}
  
where {\bf 1} is a vector containing 1 in every entry. Then, the TWSA representation at the $0.5^{\circ}$ model grid follows as
  
  \begin{equation}
    \label{eq004}
        \hat{\bf z} = {\bf U}\hat{\bf x}
    \end{equation}
    
where ${\bf U}$ is similar to ${\bf H}$ except that it accumulates the ten model storages to total storage anomaly TWSA, while applying the temporal mean value from the open loop run, on the $0.5^\circ$ model grid. Here, we again choose the period for the temporal mean similar to the temporal mean of the GRACE/-FO observation (2003-2016) because it was decided before the gap.  Similarly, the ensemble representation of GLWS TWSA at the model grid is
  
  \begin{equation}
    \label{eq005}
        \hat{\bf Z} = {\bf U}\hat{\bf X} \ .
    \end{equation}
  
From Eq. (\ref{eq005} it is straightforward to derive GLWS 1-sigma standard deviations or other error measures, including spatial correlation metrics. We note, however, that the ensemble median (for TWSA grids, $\tilde{\bf z}={\bf U}\tilde{\bf x}$ where the tilde stands for the median) is often regarded as being more representative for the variability that individual ensemble members display. The ensemble median is considered in this study when we look at extremes, i.e. total water storage anomalies during drought and flood conditions.

With GLWS2.0, we can provide in addition three more data sets that we term 'level-2' accompanied with a 1-sigma characterization; i.e. surface storage anomalies (sum of wetlands, lakes, man-made reservoirs, and river storages),  groundwater storage anomalies, and soil moisture anomalies. These are all derived via Eqs. (\ref{eq004}) and (\ref{eq005}) with the respective linear operators ${\bf U}$.

As discussed above, our GRACE/-FO error model seeks to represent coverage and orbital geometry effects (latitudinal growth of errors and anisotropic correlation) as well as the temporally varying data quality, inevitably limited by the quality of the error models used in the level-2 processing. This is complemented by an error model for the hydrological simulations that accounts for forcing data uncertainty (which depends e.g. on season and topography) and model parameter errors (which depend strongly on climate regions, soil types, and other geographical factors). As a result, the GLWS error model is expected to reflect all of these patterns and may be thus difficult to interpret. A caveat is, however, that several conceptual equations in the modeled water balances (e.g. inequality constraints whenever a variable cannot be negative, wetland flooding once a threshold storage is reached) can lead to multimodal probability distributions which are possibly not well represented through the limited-size ensemble, and for which the EnKF is known to be not optimal. In lieu of other feasible approaches (e.g. numerical costs seem prohibitive for a particle filter in our application) we accept this limitation.

\section{Characterization of the GLWS2.0 global land $0.5^\circ$ total water storage anomaly data set}
\label{sec_glws}
\subsection{Comparing GLWS2.0 to GRACE data and to WaterGAP model simulations}

In this section, we will first compare the total water storage anomaly grids derived by accumulating over all storages in the assimilation product (and removing the temporal mean) to the original GRACE data (at $0.5^\circ$ resolution), as well as to the 'open-loop' WaterGAP simulation (OL). The latter represents the mean of the free-running ensemble, assuming no data is assimilated. We will first look at linear trends, the entire signal variability expressed as RMS, and the seasonal cycle, and then add a comparison in terms of large-scale surplus and deficit water storage averaged over latitudinal bands.

\begin{figure}[H]
\centering
    \includegraphics[width=15.3cm]{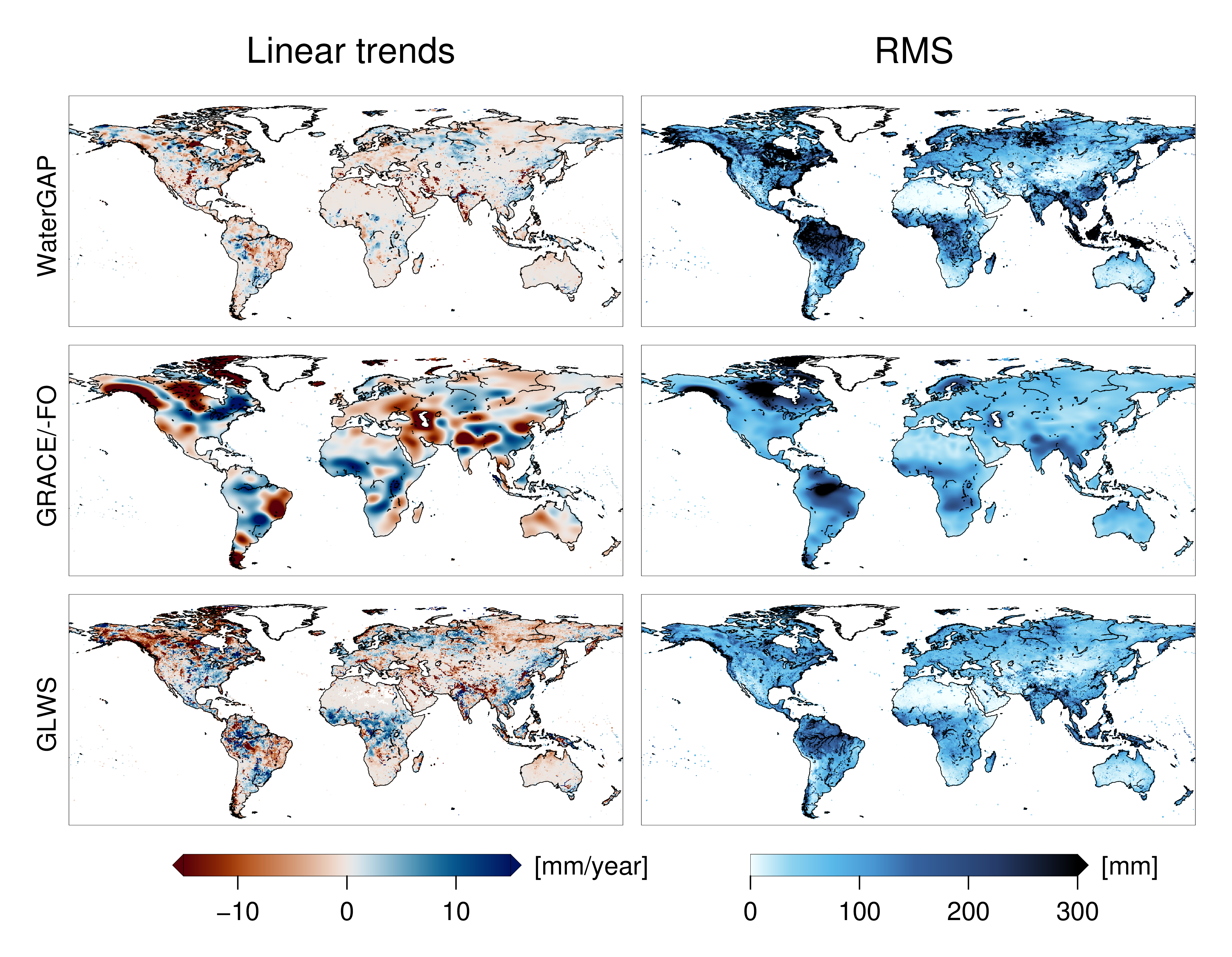}
    \caption{Linear trends (left) and root mean square variability (RMS, right) of total water storage anomalies derived by WaterGAP (top), GRACE/-FO (middle), and GRACE/-FO assimilated into WaterGAP (bottom). The linear trend and RMS are computed from 2003 to 2019. The trends are given in mm per year and the RMS variability is given in mm. }
    \label{figure_01_trends_rms}
\end{figure}

Fig. \ref{figure_01_trends_rms} shows in the left column the linear trends in total water storage anomalies, computed for years 2003-2019 from the WaterGAP model simulations (top), the original GRACE/-FO-derived TWSA data (middle) and the new GLWS2.0 data set (bottom). Root Mean Square (RMS) maps in the right column display the total variability of TWSA in the model, GRACE/-FO and GLWS2.0 data set. In the GRACE data, we identify large negative trends i.e. increasing water deficit with -15 mm/a or more, e.g. in the Parana-Sao Francisco-Atlantico river basins in South America, the Caspian Sea region, in parts of Northern India and in glacier regions such as Alaska, Patagonia, and Northern Canada. This is not surprising, as in previous studies  \citep[e.g.][]{gardner2011sharply, luthcke2013antarctica, getirana2016extreme, panda2016spatiotemporal,   chen2017long, rodell2018}, these regions were identified to have been affected by severe droughts, decreasing surface water levels, groundwater depletion and melting of glaciers. In contrast, large positive trends (i.e. wetting) can be found in Eastern Canada, West Africa, the Zambesi basin and African Great Lake region, and in the La Plata basin in South America, which were all associated with regions that are either experiencing increasing precipitation and/or transform from dry to wet conditions, e.g. recovery after a drought during the observation period takes place \citep[e.g.][]{rodell2018}.

As has been noticed in many studies  \citep[e.g.][and references therein]{scanlon_global_2018}, the WaterGAP global hydrology model generally simulates much lower TWSA long-term trends as compared to GRACE/-FO. However, we notice that there are many similarities in the signs, i.e. drying vs. wetting, e.g. in the Amazon basin, West Africa, Northwest America, and other regions. In comparison to GRACE, the model displays the trend maps with much higher spatial detail due to its intrinsically higher spatial resolution, which in turn is due to the higher spatial resolution of its forcing data. We note that drying trends in our OL mean simulations with WaterGAP2.2e appear less pronounced as in \cite{muller_schmied_global_2021} (cf. their Fig. 9 for basin-averaged total water storage trends in the 2.2d standard version).

Linear trends in GLWS seem to confirm the anticipated advantage of assimilating GRACE data into the model: The effective spatial resolution of the assimilated data set appears visually much higher than the GRACE data, while the large-scale patterns from GRACE/-FO seem inherited for many regions. We thus hypothesize that we obtain a more detailed picture of the distribution of negative and positive trends. For example, a pattern of large negative trend at the Southern tip of Patagonia that one identifies in the GRACE data appears more localized in the actual glacier location in the  GLWS data. We note that if trends for GLWS would be inferred from the ensemble median instead of the ensemble mean, this would lead to negligible average changes of about 0.02 \%.

Hydrological models are not expected to represent small trends in the water balance well \citep{scanlon_global_2018}, while many studies have shown that they compare well with GRACE/-FO data at seasonal and monthly timescales. From the right column of Fig. \ref{figure_01_trends_rms} we notice that WaterGAP seems to display variability at large spatial scales comparable to observed TWSA, with generally fewer signals in arid regions. The model cannot reproduce glacier mass changes (e.g. Alaska glacier chain) as has been pointed out already, but for many larger river basins variability is similar as seen in GRACE (as has been described in countless studies). We suspect that glacier mass loss signals missing in the model simulation will also dominate the differences in RMS. We notice that in GLWS2.0, trends in arid regions are somewhat pulled towards GRACE, and spatial detail is added in particular for large lakes, reservoirs, and riverine regions. We do see stronger variability in GLWS2.0 as compared to GRACE/-FO in island regions such as the Indonesian Archipelago, which are difficult to assess from satellite data only due to limited resolution, the need for smoothing, and the ocean mass leakage problem.

\begin{figure}[H]
\centering
    \includegraphics[width=15.3cm]{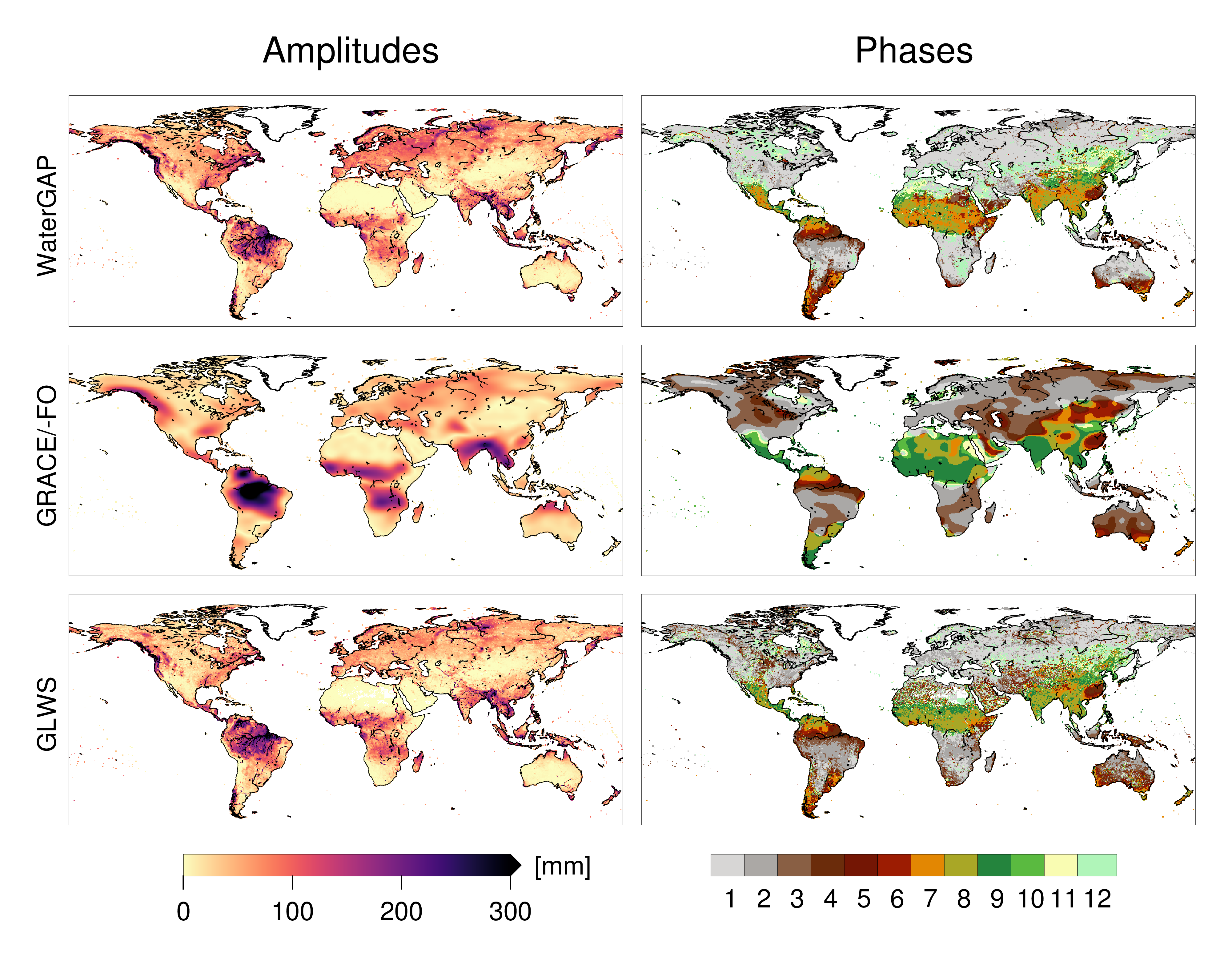}
    \caption{Annual amplitudes and phases of total water storage anomalies derived by WaterGAP (top), GRACE/-FO (middle), and GLWS2.0. The amplitudes are computed for the 2003 to 2019 time frame and given in mm. The phases are computed for the same period and indicate in which month the maximum amplitude occurs (1 = January).}
    \label{figure_02_amplitudes_cosine_sine}
\end{figure}

Next, the dominating annual signals in all three data sets are compared in Fig. \ref{figure_02_amplitudes_cosine_sine}, with the TWSA amplitude in the left column and the phase (month in which the maximum amplitude occurs) in the right column. Since in most river basins the signal variability is dominated by an annual signal, the left column is indeed very similar to the right column in Fig. \ref{figure_01_trends_rms} (this is not true for large reservoirs where long-term water level changes due to resource management dominate). More interesting is the right column, which reveals that phase mismatches of one or two months between GRACE/-FO data and model simulation are not uncommon. While these are (partly) mitigated through assimilation in some regions (West Africa, Central America, Australia, South Asia), one-month mismatches remain between GLWS2.0 and GRACE/-FO phases e.g. for Siberia or large parts of North America. We note, however, that GLWS inherits the spatial variability of phase (e.g. regional variations in the onset of the rainfall season) from WaterGAP which is probably real, and which cannot be represented in GRACE/-FO data due to coarse effective resolution. For the sake of completeness, we note that if we would determine the GLWS annual amplitudes from the ensemble median instead of the ensemble mean, differences would average to insignificant 0.13 \%.
 
\begin{figure}
\centering
    \includegraphics[width=17.3cm]{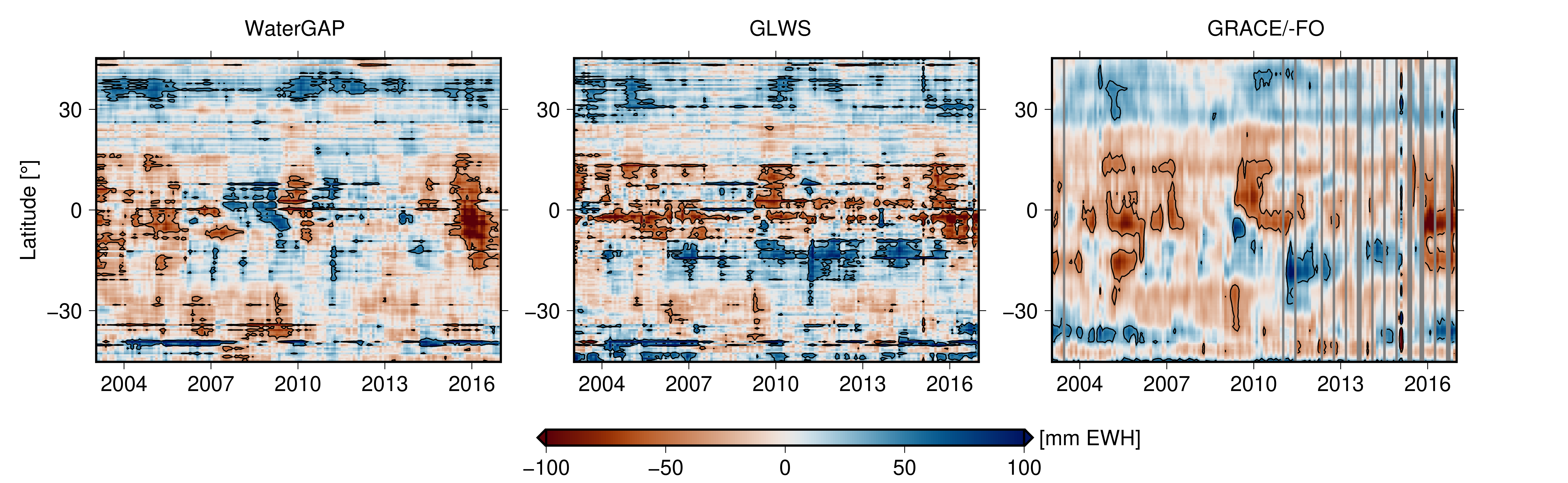}
    \caption{TWSA averaged across longitudes for open-loop simulation, GLWS, and GRACE/-FO from 2003 to 2019 shown as Hovmöller diagram. The linear trend, constant bias, and annual and semi-annual signal were removed to derive nonseasonal TWSA. Here we use the ensemble median of the water states.}
    \label{figure_03_hovmoeller_tapley_median}
\end{figure}

We now add the temporal dimension to these comparisons and ask to what extent extreme water storage anomalies are represented in GLWS2.0. To this extent, in the right column of Fig. \ref{figure_03_hovmoeller_tapley_median}, the Hovmöller diagram for TWSA as observed by GRACE/-FO is shown. This figure (we borrow the representation from \cite{tapley_contributions_2019}) represents storage anomalies with respect to the mean, trend, and seasonal cycle, which have been attributed to drought events in the Amazon, southern Africa, and Australia in 2005, exceptional flooding in the Amazon basin and in south Central Africa in 2009, ENSO-related precipitation in Australia, Amazon, and southern Africa in 2011, and droughts across the Amazon, southern Africa including Madagascar, and Australia in 2015/2016. Our satellite TWSA diagram differs slightly from \cite{tapley_contributions_2019}; this might be due to the use of a different GRACE solution but we cannot exclude that our method for removal of the seasonal cycle differs from theirs.

All major extreme events that had been identified in the GRACE data appear present in GLWS2.0 (as noted earlier, we show the ensemble median here). We notice that, strikingly, the La Nina-related wet period  2007-2012 \citep{sori_oceanic_2021} appears more pronounced and longer-lasting in GLWS2.0 for southern latitudes $-10^\circ$ to $-20^\circ$ (where anomalies in the OL seem to fit better to GRACE) whereas the unprecedented 2016 drought in South America \citep{jimenez-munoz_record-breaking_2016, erfanian_unprecedented_2017} appears less pronounced south of about $-10^\circ$ latitude. We find that in general tropical drought events become visible much more spatially localized as compared to GRACE, which is expected and generally corresponds to precipitation-derived drought indicator maps. GLWS2.0 inherits many drought and flood events from the model simulation, with mostly wet conditions more pronounced. As a counter-example, the 2010 drought-related  water storage deficit (e.g. Amazon, Horn of Africa, China, U.S.A., Mexico) extends somewhat further to northern latitudes in GLWS as compared to the simulation median. However, a caveat is that the model open loop median differs from the WaterGAP standard run, and thus such differences should not be over-interpreted. It is also important to remember that anomalies in the maps refer, by definition, to the individual mean, trend, and annual behavior, which is not necessarily identical across the three data sets. The reader should also be reminded that GRACE-derived TWSA extremes are far from being Gaussian distributed, with significant skewness in their distribution \citep{forootan_separation_2012}.

\subsection{Comparing GLWS2.0 to GRACE data and to WaterGAP model in the spectral domain}
In this part, we seek to compare the assimilation data set with GRACE and model data in terms of spherical harmonic representation. With what we have seen in the previous section, it is expected that GLWS inherits spatial resolution from WaterGAP while at large spatial scales there seems somewhat less variability present as compared to GRACE/-FO (e.g. Fig. \ref{figure_02_amplitudes_cosine_sine}. For a consistent and systematic assessment, we thus map global land GLWS, WaterGAP, and also GRACE/-FO total water storage maps to spherical harmonics. It is clear that the 'GRACE/-FO TWSA harmonics' derived in this way do not necessarily resemble the GRACE solutions, as they represent (DDK-)filtered solutions masked to the continents only and not including contributions from Greenland and Antarctica ice sheets (global land area $\cal L$, Load Love number $k'_n$, TWSA in equivalent water height $\Delta h$):

    \begin{equation}
    \label{eq020}
    c_{nm} = \frac{3\rho_w}{\rho_e} \frac{1+k'_n}{2n+1} 
    \int_{\cal L} C_{nm} \ \Delta h \ d\omega \qquad
    s_{nm} = \frac{3\rho_w}{\rho_e} \frac{1+k'_n}{2n+1} 
    \int_{\cal L} S_{nm} \ \Delta h \ d\omega \ .
    \end{equation}

\begin{figure}[H]
\centering
    \includegraphics[width=15.3cm]{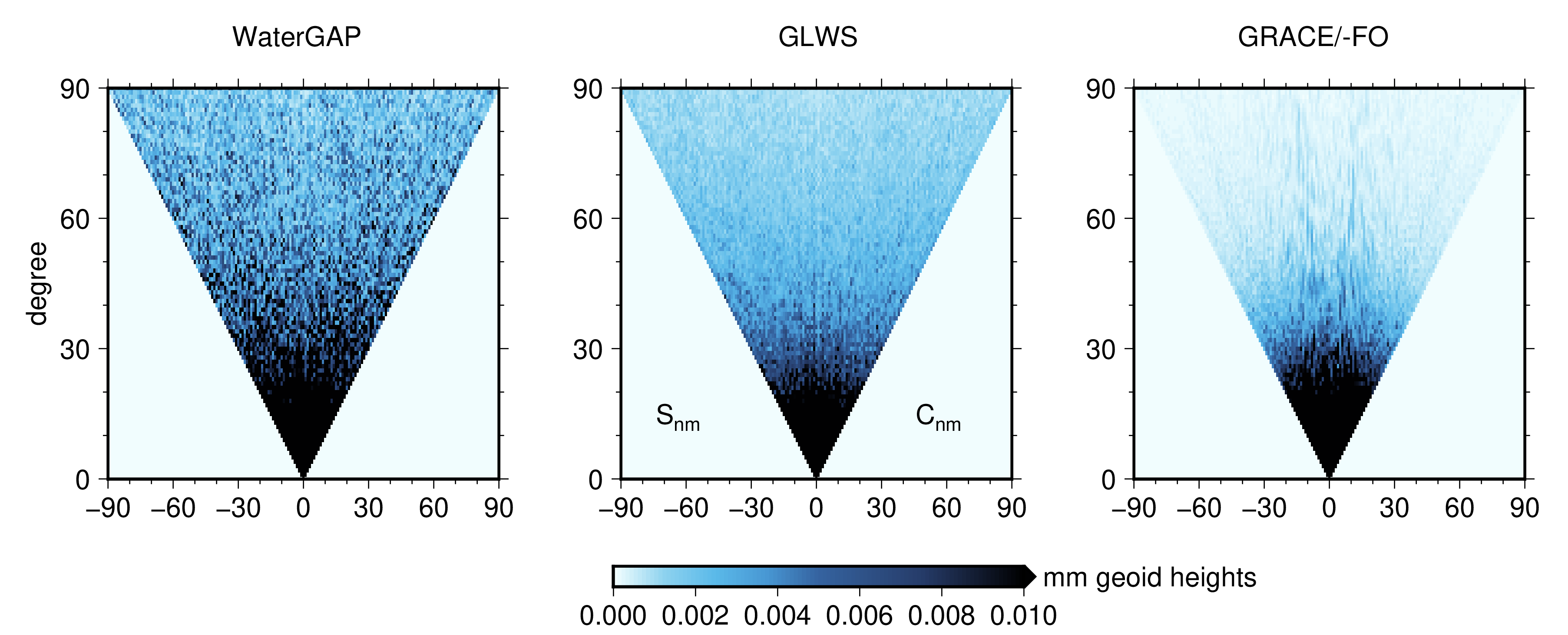}
    \caption{RMS of spherical harmonic coefficients derived by WaterGAP (left), GRACE/-FO (right), and GLWS2.0 (GRACE/-FO assimilated into WaterGAP for 2003 to 2019, middle). The RMS is expressed in terms of geoid variability here, in mm-units.}
    \label{figure_04_meanSHC}
\end{figure}

Fig. \ref{figure_04_meanSHC} shows the monthly RMS of the three time series per coefficient. For ease of comparison, we remapped all triangle plots such that colors directly signify the contribution to geoid variability, in units of mm. We notice that the DA indeed effectively removes much high-frequency content from WaterGAP, beyond about degree 25, but it retains significantly more variability as compared to GRACE/-FO beyond about degree 35.
    
\begin{figure}[H]
\centering
    \includegraphics[width=14.3cm]{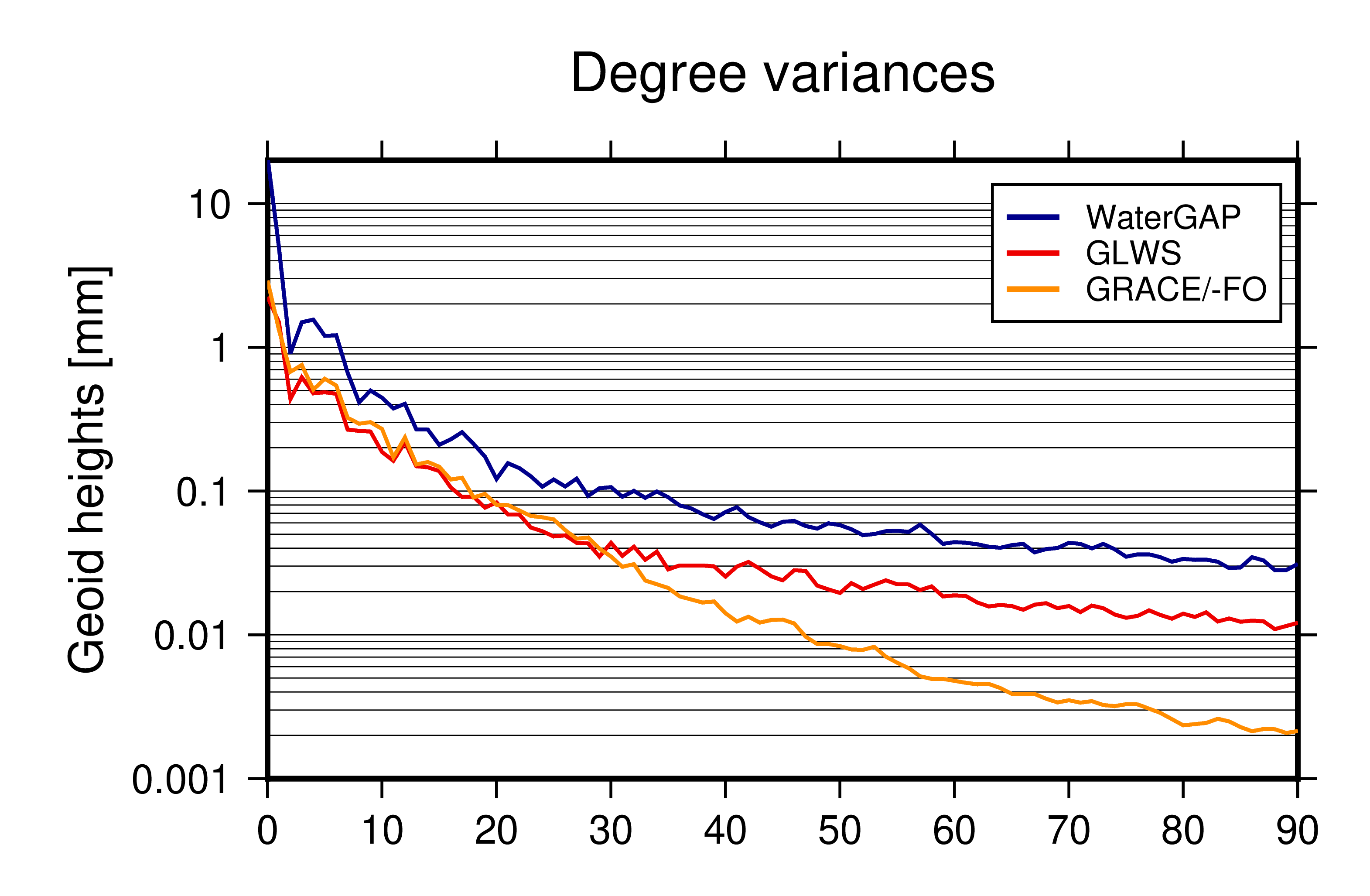}
    \caption{Degree variance for GRACE, WaterGAP, and GLWS in August 2010 expressed in geoid heights [mm].}
    \label{figure_05_varSHC}
\end{figure}

The same can be observed when the triangle plots are collapsed to degree variance curves, as in Fig. \ref{figure_05_varSHC}. We notice a striking offset between WaterGAP and GRACE/-FO which we speculate could be caused by the much larger signal over Siberia and North America seen in the hydrological model. The GLWS data is much closer to GRACE/-FO at low degrees, whereas at around degree 30 it is evidently pulled towards exhibiting more power. Here, GLWS seems to adjust to the same power decay law as seen in WaterGAP.

\begin{figure}[H]
\centering
    \includegraphics[width=10.3cm]{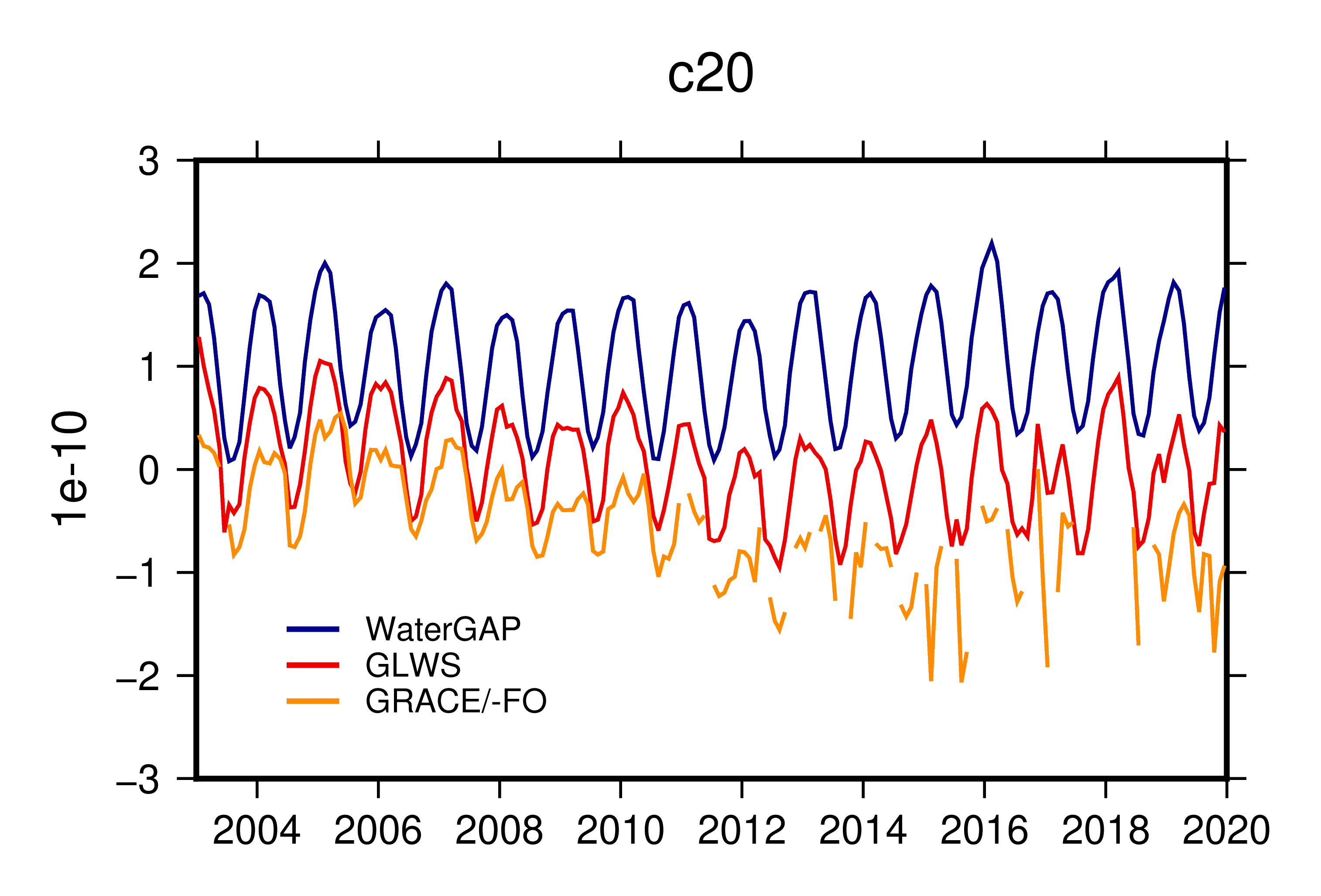}
    \caption{$c_{20}$ spherical harmonic coefficient derived by WaterGAP (top), GRACE/-FO and GLWS2.0 (GRACE/-FO assimilated into WaterGAP computed for 2003 to 2019).}
    \label{figure_06_c20SHC}
\end{figure}

It is instructive to view how single spherical harmonic coefficients get modified through assimilation. We chose $c_{20}$ to be displayed in Fig. \ref{figure_06_c20SHC}; again we remind the reader that the 'GRACE coefficient' time series shown here is (1) derived from a GRACE/-FO solution where the original $c_{20}$ has been replaced by an SLR solution, and (2) TWSA had been synthesized on the global land and remapped into the spherical harmonic domain. The WaterGAP open loop dynamic flattening contribution appears with much less interannual variability, yet it shows a larger annual amplitude with less within-year variation. GLWS2.0 $c_{20}$ inherits properties from both data sets; it remains to be understood what is closer to reality. Gaps in the GRACE/-FO time series are reasonably filled, but the GLWS2.0 $c_{20}$ seems to inherit stronger fluctuations from the post-2010 GRACE/-FO data. The annual amplitude seems slightly reduced as compared to WaterGAP. We notice that GLWS seems to follow GRACE/-FO closer (but with a bigger amplitude) before about 2011; this is likely an effect of the less accurate GRACE solutions after this date being reflected in our error model.

\subsection{Uncertainty of GLWS TWSA derived from the ensemble}
As described in the previous section, uncertainties in the GLWS2.0 assimilation data set are represented through the ensemble spread ${\bf W}=\hat{\bf X}-{\bf 1}\hat{\bf x}$, which depends on the implemented GRACE TWSA error model ($\bf G$) as well as on the error models that we implemented with the WaterGAP simulation, the boundary conditions (e.g. soil and vegetation maps represented in the model) and the actual model state (e.g seasonal wetness). As can be expected, in the absence of GRACE/-FO data, i.e. during monthly gaps and during the gap between GRACE and GRACE-FO, the error in GLWS2.0 grows faster as compared to the early GRACE phase. Averaged month-to-month, $0.5^\circ$ cell TWSA uncertainties are found as between about 2 cm in arid regions and can reach more than 15 cm in the Amazon during GRACE/-FO data gaps.
 
In Fig. \ref{figure_07_trends_std} below, we show, as an example, the propagated uncertainty of the 2003-2019 linear total water storage trends in GLWS2.0. We notice that indeed for many regions that exhibit large trends also larger uncertainties prevail, but these uncertainties (that refer to $0.5^\circ$ grid cells) are generally below 1mm/a except in some cells where large surface water bodies are present. We notice that trend uncertainties are particularly low in large arid regions, which is likely due to confidence in GRACE/-FO data but is certainly also related to an underestimation of model representation errors in our WaterGAP ensemble.

\begin{figure}[H]
\centering
    \includegraphics[width=10.3cm]{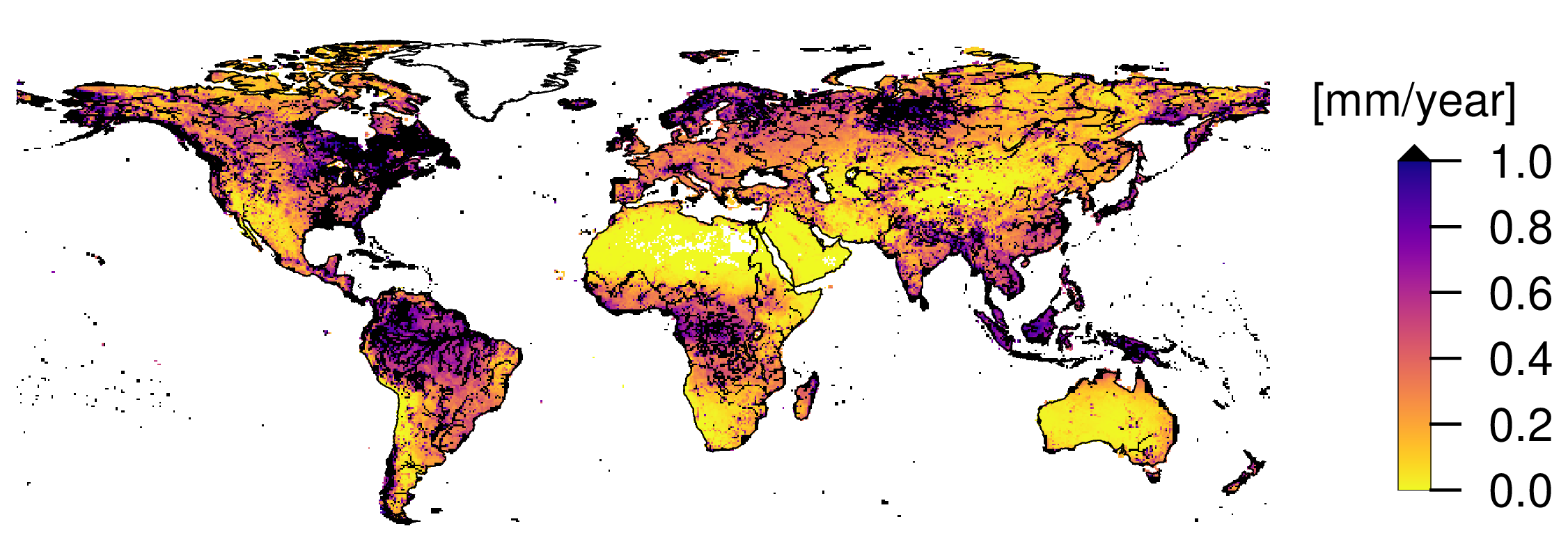}
    \caption{GLWS standard deviations of linear trends computed from 2003 to 2019. The standard deviations are given in mm per year.}
    \label{figure_07_trends_std}
\end{figure}

A caveat is that our GRACE/-FO error model does account for the non-diagonal error structure of the spherical harmonic coefficients due to orbital patterns and measurement principle, and thus to the dominating North-South dependence of errors, but it disregards the resulting spatial correlation at the grid cell level. In other words, one should not use the ensemble to represent spatial correlations between GLWS TWSA across grid point locations.

A second concern must be whether the GRACE/-FO uncertainty e.g. due to insufficient background modeling in the processing can be adequately represented at all through such propagated errors; however, since all published GRACE/-FO level-2 solutions basically utilize the same processing standards and background models we do not see another way of representing GRACE/-FO errors more realistic. A third issue could be that the limited ensemble error representation in the EnKF may lead to additional artifacts. Square root ensemble filters have been developed to somewhat mitigate the latter problem and future GLWS versions may benefit from these implementations.

When referring to Eqs. (\ref{eq002}) and (\ref{eq003}), it becomes obvious that the best estimate in the EnKF is constructed in the 'observation space', i.e. as a weighted linear combination of the observation ensemble ${\bf Y}$ (which, in our case, consists of 'clones' of a single GRACE/-FO solution  perturbed with errors sampled from the covariance matrices provided by ITSG, TU Graz). This points to an interesting analogy with current efforts (e.g. within the COST-G project) to find optimal weights for combining different GRACE/-FO solutions.

\subsection{Robustness of the assimilation against analysis choices}
\label{sec_robustness}
Our data assimilation framework allows one to choose certain 'tuning' parameters. To understand how robust the GLWS2.0 maps are with respect to these choices, we have first rerun the assimilation while varying the inflation factor (cf. Tab. \ref{table_DA_settings}; inflation increases the ensemble spread, to avoid filter 'saturation' and a decreasing impact of observations). However by increasing this factor by 25 $\%$ we derive a globally averaged RMS change of only 3 $\%$ when compared to GLWS2.0. We suggest here that moderate inflation has little impact when being weighted against the 1-sigma observation error. For the next analysis, we return to our previous settings and we replace the 1-sigma TWSA errors by 2-sigma (3-sigma) -- this leads to a change of about 9.6 \% ( 11.6 \%) in RMS globally averaged. These differences are still relatively minor. The situation is slightly different when we modify the spatial resolution of the input GRACE/-FO TWSA grids for the assimilation (data matrix ${\bf Y}$). When moving from 4$^\circ$ to 5$^\circ$ grid cells, we find that the assimilation TWSA differs by 20 \% RMS as compared to GLWS2.0. However, this is not unexpected since with about 50 \% bigger input cell area, the coarser grid will be less able to represent real spatial variability of hydrological regimes. On the other hand, the coarser grid contains less observation and thus the weighting between model and data is additionally shifted towards the model. Last, if we replace the GRACE/-FO solution with solutions from CSR or GFZ, the RMS globally averaged differs by 11.3 $\%$ or 11.5 $\%$  (16.5 $\%$ or 21.7 $\%$ prior to assimilation), respectively, thus, the choice of solutions has a lower impact than the choice of grid size.

\subsection{Comparisons with other global TWSA data sets}
In the following, we compare GLWS2.0 TWSA  against another global land storage data set, which was developed by assimilating GRACE/-FO mascon data provided by NASA's Goddard Space Flight Center (GSFC) into the Catchment Land Surface Model \citep[CLSM,][]{koster_catchment-based_2000}, a widely used global land surface model on 36 km spatial resolution \citep{girotto_assimilation_2016,li_global_2019,felsberg_global_2021}.

\begin{figure}[H]
\centering
    \includegraphics[width=15.3cm]{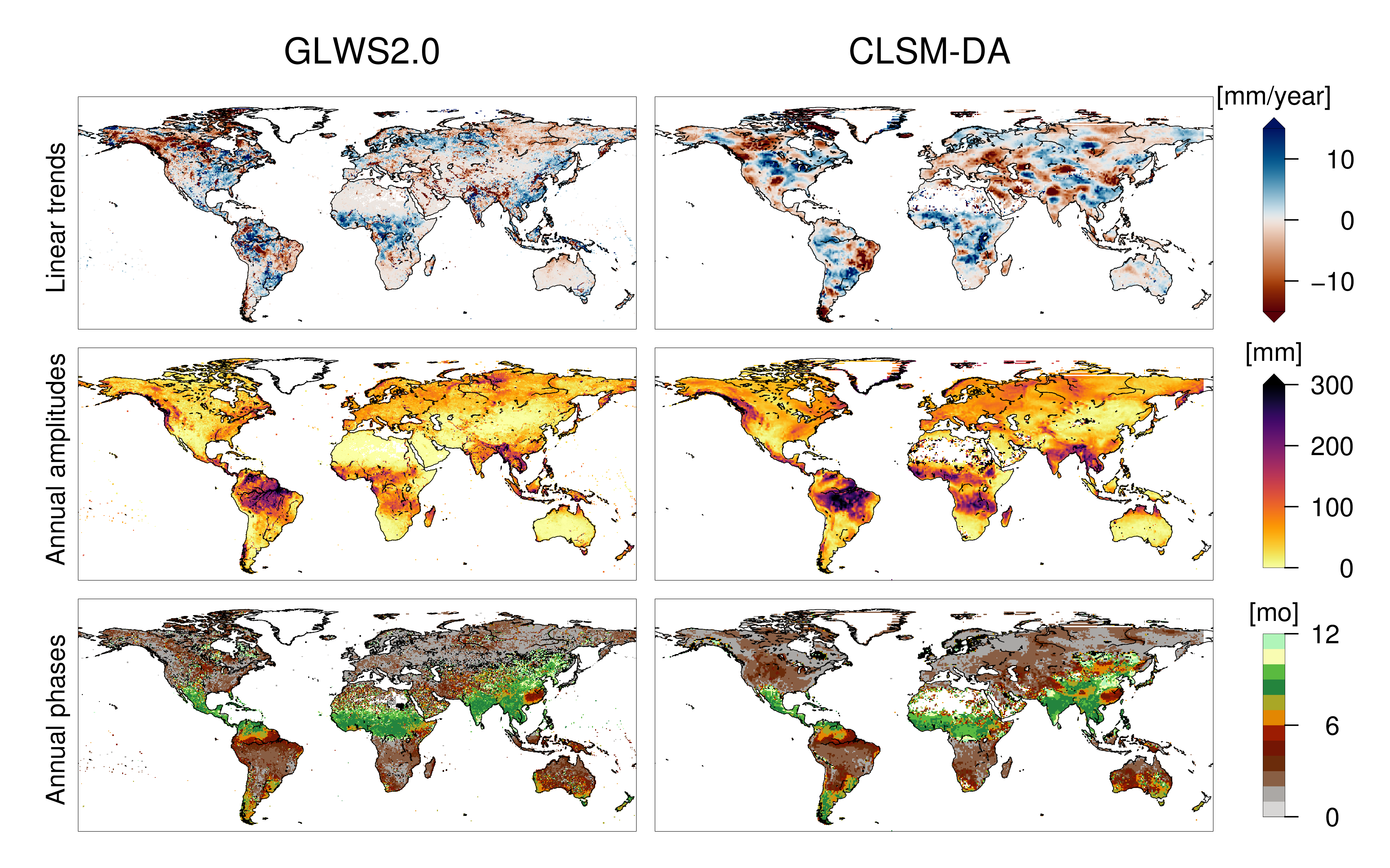}
    \caption{Comparison of the two $0.5^\circ$ GLWS2.0 and CLSM-DA TWSA data sets in terms of trends [mm per year], annual amplitudes [mm], and phases of the annual amplitudes [months]. The data period is 2003-2019.}
    \label{figure_08_clsm_glws_compared}
\end{figure}

Overall, we identify a good agreement between both assimilation data sets (Fig. \ref{figure_08_clsm_glws_compared}). Important, we find linear trends and annual amplitudes (determined via multi-linear regression as in \cite{gerdener2022}) in the two assimilations closer to the respective GRACE/-FO products than the model simulations (in both cases the ensemble mean); in other words, integrating GRACE/-FO data into the model simulations by assimilation results in trends and seasonal cycle converging. However, when compared to the CLSM-DA, GLWS2.0 exhibits somewhat lower trends and lower annual amplitudes. For example, for CLSM-DA the annual amplitudes in the Zambezi river basin are up to 200 mm, while with GLWS2.0 we find amplitudes reaching about 100 mm in this basin. This difference is mainly confirmed for humid areas, while also for some arid areas like the Sahelian zone or Australia’s interior minor differences in spatial patterns of trends and amplitudes can be observed.

We suggest that differences in the two DA products are mainly due to differences in data and model representation. Whereas GLWS2.0 assimilates gridded GRACE/-FO TWSA maps derived from (DDK3-) filtered spherical harmonic coefficients, the CLSM-DA ingests GRACE/-FO mascons and applies a rescaling of the observations before assimilation. Moreover, CLSM and WaterGAP models differ in concepts and forcing data. For example, the WaterGAP model has an explicit representation of water stored in river networks and adjacent floodplains, and therefore for large rivers, we find the river routing pattern imprinted in the GLWS2.0, which is not the case for the CLSM-DA. WaterGAP uses a single soil moisture storage and a groundwater storage compartment per grid cell, whereas in CLSM the soil moisture and groundwater are diagnosed from three prognostic variables that have an assumed within-grid cell spatial distribution. Lastly, CLSM-DA uses MERRA2 forcing data while WaterGAP is forced with GSWP3-W5E5. We suggest that future studies should address the effect of assimilation system design and tuning choices across multiple models, with consistent GRACE/-FO input grids and climate forcing data.

\section{Validation with GNSS}
\label{sec_validation}
As was mentioned above, GNSS time series are hypothesized to sense the redistribution of total water storage through elastic crustal loading, provided that all other relevant processes can be modeled and removed. We use 1030 detrended vertical displacement time series that were provided by the IGS within the Repro3 reprocessing (Rebischung, 2021).
 
\begin{figure}[H]
\centering
    \includegraphics[width=15.3cm]{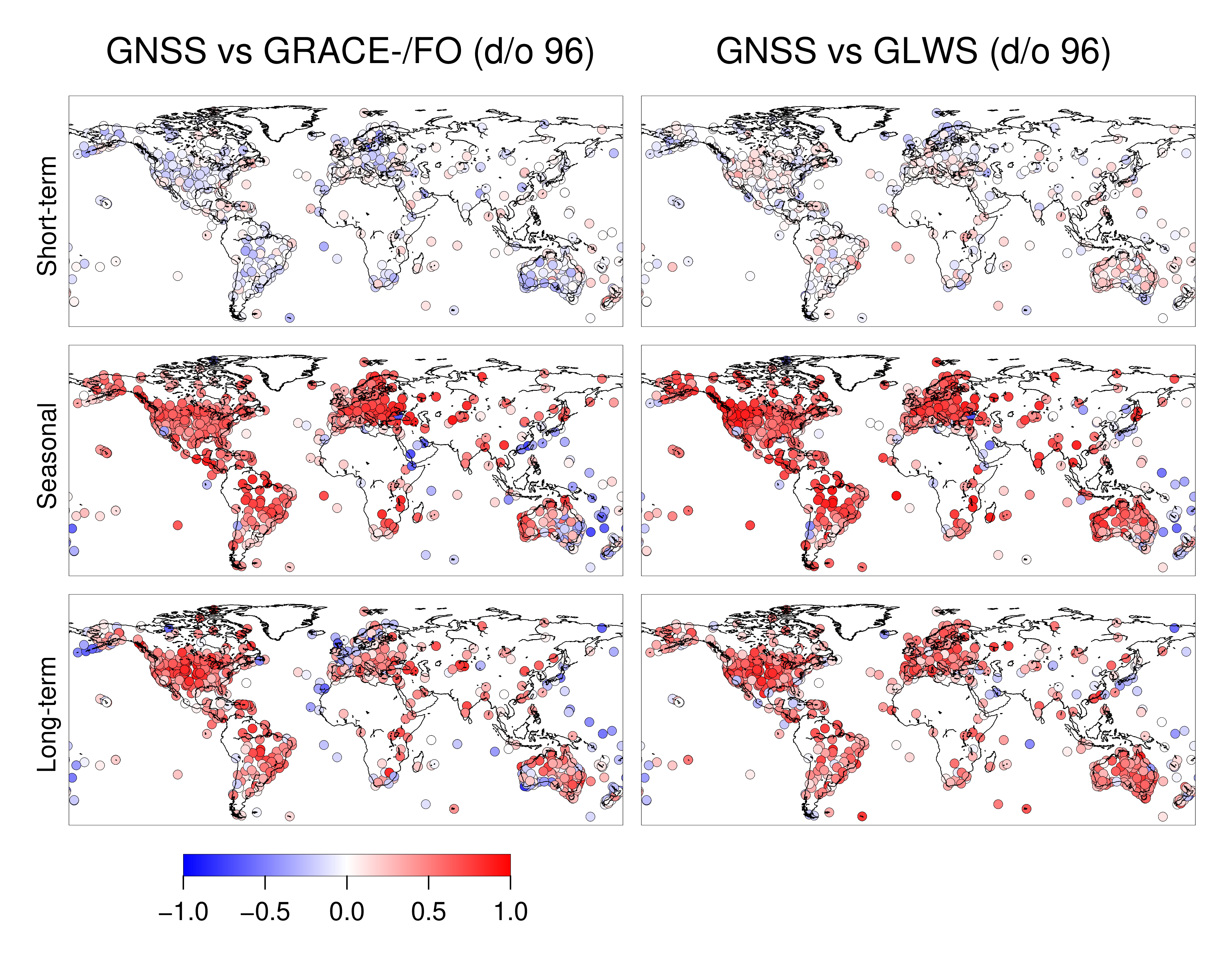}
    \caption{Comparison of correlation coefficients for short-term, seasonal, and long-term temporal bands, GNSS vs. GRACE/-FO (left) and GNSS vs. GLWS2.0 d/o 96 (right), all 2003 to 2019.}
    \label{figure_09_GNSS_corr_maps}
\end{figure}

Fig. \ref{figure_09_GNSS_corr_maps} shows a direct comparison of predicted monthly vertical motion time series from GRACE/-FO and GLWS2.0, in terms of correlation. We split all residual time series, i.e. the unexplained variability, into three temporal bands, following a wavelet decomposition approach \citep{meyer_1990, Bogusz2013}. After partial reconstruction, the short-term band contains all temporal wavelengths from 2 months to 5 months, the seasonal band includes the signal between 5 months and 1.4 years, and the long-term band represents the wavelengths beyond 1.4 years.
 GLWS2.0 TWSA maps were transformed to the spectral domain, in order to facilitate the loading computation similar as Eq. (\ref{eq020}) with the load Love number involving $h'_n$.

Here we truncated the assimilation TWSA first at d/o 96, i.e. spectrally consistent with GRACE. For GLWS2.0, we note some improvement at shorter timescales in particular in Europe but also in North America, Southern Australia, and South America as compared to GRACE/-FO. At seasonal timescales, improvement of GLWS2.0 w.r.t. GRACE/-FO data is found most notably over Australia and the US and Canada West coast. At longer than seasonal timescales, we find a noticeable improvement over Western and Central Europe and Australia, whereas it appears that for some stations in South America and Northern Canada the original GRACE/-FO data show a better correlation. 

\begin{figure}[H]
\centering
    \includegraphics[width=15.3cm]{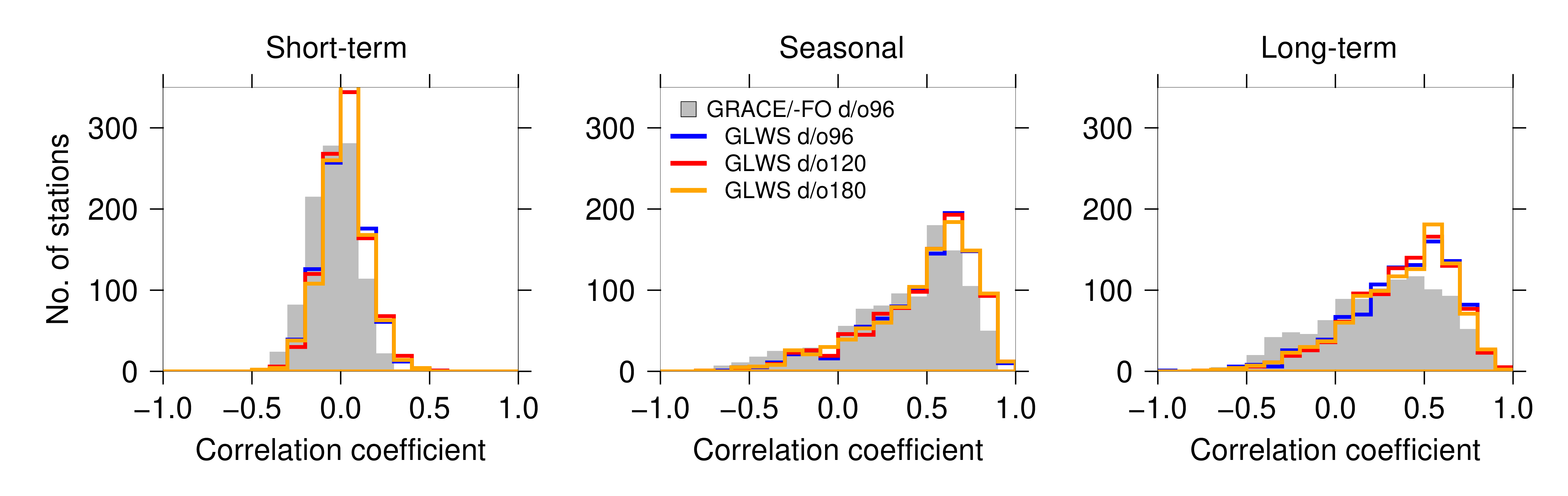}
    \caption{Global statistics of correlation  for different temporal bands, GRACE/-FO vs. GLWS2.0 d/o 96, all 2003 to 2019.}
    \label{figure_10_GNSS_corr_stats}
\end{figure}

All in all (Fig. \ref{figure_10_GNSS_corr_stats}), we find a better agreement in all three temporal bands for the GLWS2.0 total water storage grids as compared to GRACE/-FO. We note this result is not related to the original $0.5^\circ$ spatial resolution in GLWS2.0. It is evident, however, that all statistics are leaning heavily towards Europe, North America, and Australia due to the high station density in these regions and should not be interpreted as being globally representative. Future studies should focus on removing this geographical bias, e.g. by thinning out the GNSS sites for comparison.

The spectral resolution of d/o 96 corresponds to about 200 kilometers spatial wavelength. When we truncate GLWS2.0 spherical harmonic synthesis at d/o 120 or 180, slight but consistent improvements are observed in all three temporal bands (red and orange bars in Fig. \ref{figure_10_GNSS_corr_stats}); yet these are minor when compared to the remaining variability between GRACE and GLWS. While this may seem counter-intuitive, we note that even d/o 180 corresponds to about 110 km spatial resolution and will lack many real local hydrological signals due to missing resolution, but foremost since the elastic crustal loading smoothes out effectively such effects. This is also consistent with what \cite{springer_evidence_2019} observed in experiments with an $0.08^\circ$ resolution hydrological model. In summary, comparisons of predicted non-linear hydrological loading to Repro3 GNSS time series suggest that GLWS seems to fit better at all temporal bands than GRACE/-FO. Given that some noise floor must be expected in these comparisons due to the presence of real non-loading, e.g. tectonic displacements and remaining GNSS technique errors \citep{chanard_warning_2020}, this is not necessarily an expected result.

\section{Conclusion}
We describe a new $0.5^\circ$ global land water storage data set, GLWS2.0, which is derived via assimilating maps of GRACE/-FO TWSA into a hydrological model, WaterGAP. Although GLWS2.0 knows all storages that WaterGAP represents, i.e. soil moisture, various surface water body storages, snow, and groundwater storage, our focus here is on the TWSA representation in GLWS2.0.

We find that GLWS2.0 matches GRACE/-FO TWSA in terms of trend regions and variability, with trends being generally less pronounced, likely inherited from the WaterGAP open-loop runs and possibly too low. GLWS2.0 displays less variability in arid regions as compared to GRACE/-FO. Much more spatial detail is present in particular in trend regions. Although GRACE data and the hydrological model differ, GLWS2.0 shows many similarities with the CLSM-DA assimilation. In the spectral domain, GLWS2.0 exhibits more power as compared to GRACE/-FO from about degrees 25 onwards, reaching nearly the same slope of the spectrum as the WaterGAP open-loop ensemble mean. This is, to our knowledge, the first assessment of the effect of GRACE/-FO assimilation in the spectral domain. In terms of extreme water storage anomalies, GLWS2.0 reveals that many droughts and flooding events observed with GRACE/-FO indeed exhibit more spatial detail.  Beyond this - in fact expected - result, the assimilation suggests that in particular southern-hemisphere precipitation related water storage surplus was temporally more sustained than in GRACE solutions.

Comparisons of predicted non-linear vertical station motions to coordinate time series derived within the Repro3 contribution to ITRF2020 indicate that GLWS2.0 seems to fit better to independent GNSS observations at all temporal bands (except the linear trends which need to be removed for this kind of comparison), even when truncated to the same spectral degree as GRACE/-FO. 
It is tempting to ascribe this to the contribution of the hydrological modeling at shorter spatial scales (beyond about d/o 30 in the spectral domain), but we feel it would be premature for such conclusions. If true it would suggest that merging the GRACE/-FO data with a hydrological model would generally improve TWSA representation. It would also support the idea to integrate GNSS-derived loading time series in assimilation studies. We anticipate that more validation experiments and a generally improved understanding of the loading signals observed in GNSS are required here.

We believe that GLWS2.0 provides a useful data set of total water storage with true $0.5^\circ$ resolution, in the sense that at shorter spatial scales gradually the information content from the hydrological model, and thus implicitly its' forcing data, starts to dominate. We suggest that GLWS2.0 could be used for geodetic and geophysical applications like e.g. loading predictions for geometric space techniques, hydrological angular momentum computations, or  inversions for rheological properties. Companion studies will evaluate its applications in hydrology and climate research.

What is missing so far is an assessment of the daily TWSA as represented in the assimilation. Unlike e.g. in \cite{girotto_assimilation_2016}, TWSA is updated in our framework at the end of a month, and a meaningful interpretation of daily model states may require that we implement a temporal smoothing of the update (e.g. three times per month as in \cite{zaitchik_assimilation_2008}). Future versions of GLWS will also integrate other observables such as streamflow or snow cover fraction, primarily with the intention to improve the disaggregation of storages and the realism of the simulation of these variables. It is possible but not guaranteed that this will improve TWSA representation.  Forthcoming work will, in addition, address the estimation of model parameters within the EnKF, which is expected to improve model prediction skills e.g. during GRACE/-FO data gaps, and the skill of the assimilation (and thus future versions of GLWS) when we use different filter algorithms, for example, local filters. We will also work toward better understanding the uncertainty of GLWS2.0, and improving its representation within the ensemble.

\renewcommand\thefigure{\thesection.\arabic{figure}}    
\appendix
\setcounter{figure}{0}
\section*{Statements and declarations}   


\textit{Author Contribution Statement.}
JK conceived the presented research with support from PD. HG, JK, and AK planned the experiments. HG and KS carried out the implementation and HG performed the computations. AK contributed the GNSS validations. JK and HG wrote the manuscript with input from all authors.

\textit{Competing interests.} The authors declare that they have no known competing financial interests or personal relationships that could have appeared to influence the work reported in this paper.

\textit{Data Availability Statement.}
The GLWS2.0 data described in this contribution can be downloaded at PANGAEA: \\
Gerdener, H., Schulze, K., Kusche, J. (submitted): GLWS 2.0: A 0.5° Global Land Water Storage data set from assimilating GRACE and GRACE-FO into the WaterGAP global hydrology model from 2003-01 to 2019-12 (Level 3).

\textit{Acknowledgments.}
We acknowledge Olga Gassen (Engels) for all her effort and contribution to the data assimilation framework and Hannes Müller Schmied and Sebastian Ackermann for providing the WaterGAP model and improving its interfaces for the data assimilation framework. We thank Makan Karegar, Christian Mielke, Bell Kuei-Hua Hsu, Tonie van Dam, Anita Thea Saraswati, Laura Jensen, and Annette Eicker for providing feedback that helped us to improve the GLWS processing. We are grateful to Gabrielle De Lannoy and Manuela Girotto for providing CLSM-DA data and for the discussions.

\textit{Funding.}
This work was supported by the German Federal Ministry of Education and Research (BMBF) in the framework of the 'GlobeDrought' project (grant number 02WGR1457A) and the German Research Foundation (DFG) in the framework of the Research Unit 'Understanding the global freshwater system by combining geodetic and remote sensing information with modeling using a calibration/data assimilation approach' (GlobalCDA, grant number KU1207/26-1) and the SFB 1502/1–2022 - grant number 450058266.

\bibliography{Gerdener_2022_Global_land_water_storage.bib} 
\bibliographystyle{apalike-ejor}

\end{document}